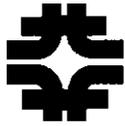 **Fermi National Accelerator Laboratory**

# A Virtual Library of Technical Publications

The Technical Publications Working Group
Elizabeth Anderson, Robert Atkinson, Elizabeth Buckley-Geer, Cynthia Crego, Lisa Giacchetti,
Stephen Hanson, David Ritchie, Jean Slisz, Sara Tompson and Stephen Wolbers

*Fermi National Accelerator Laboratory*
*P.O. Box 500, Batavia, Illinois 60510*

October 1997

# A Virtual Library of Technical Publications
## The Fermilab Technical Publications Fileserver Project
## Pilot and Phase I Report

By the Technical Publications Working Group

October 1997

## Introduction

Through a collaborative effort, the Fermilab Information Resources Department and Computing Division have created a "virtual library" of technical publications that provides public access to electronic full-text documents. This paper will discuss the vision, planning and milestones of the project, as well as the hardware, software and interdepartmental cooperation components.

## Driving Forces

In 1993 the head of the Publications Office and the Library administrator proposed to the head of Laboratory Services Section (LSS) a project to make Fermilab full-text scientific and technical publications available on the Internet. The project proposal was in part driven by the DOE/CENDI Electronic Exchange Initiative and the 1993 DOE Electronic Exchange of Scientific and Technical Information Strategic Plan. (The goal identified in the strategic plan was to make electronic exchange of full-text DOE scientific and technical information the norm by the year 2000.) However, the major force driving this project was customer demand. The physics community, which the Publications Office and the Library serve, wanted faster, more convenient access to scientific information. From the time the project was proposed, the vision of the working group has been to make the full-text of scientific and technical information developed at Fermilab fully accessible from every researcher's desktop.

## Reorganization and Cooperation

The head of Laboratory Services Section approved the proposal and also the request to seek project assistance from the Computing Division. He then submitted the proposal to the Directorate along with a request for equipment funding. After the proposal was approved by the Directorate and equipment funds were earmarked, the proposal was then sent to the Computing Division. The deputy head of the Computing Division arranged two meetings with the head of the Publications Office, the Library administrator and members of the Operating Systems Support Department to discuss the proposal and the level of support sought from the Computing Division. She agreed to provide personnel to assist LSS with writing the hardware specifications and completing the requisitioning process for the purchase of a technical publications fileserver. She also agreed to appoint Computing Division staff to serve with Laboratory Services Section staff on a Technical Publications Fileserver Working Group.

The appointed working group was comprised of members of the Publications Office, the Library, and the Computing Division's Operating Systems Support Department, Databases and Information Department, and the CDF Computing and Analysis Group. The group members included information specialists, computing specialists and physicists. Their initial charge was to study the feasibility of full-text electronic exchange of technical reports and establish support parameters. After conducting a feasibility study that showed the project to be viable, the working group was then asked to refine the project goals, identify the overall project strategy, develop an initial project plan and implement a pilot program.

Shortly after the working group was convened, the Library and the Publications Office merged to form the Information Resources Department. The purpose of the reorganization was to consolidate resources to enhance information management and facilitate the electronic exchange project.

Over the next several months, the working group laid the groundwork for the fileserver project which included formalizing with a Memorandum of Understanding (MOU) the partnership between the Computing Division and Information Resources Department, making hardware recommendations, securing earmarked equipment funds, and developing an acquisition and implementation plan.

Under the MOU, the Computing Division agreed to provide technical support for the project, which included setting up the server and peripheral equipment, housing the server, and providing system backups and upgrades. The Information Resources Department agreed to provide primary system administration of the server, workflow management, information organization, and customer service and training.

**Project Implementation**
The fileserver system hardware was purchased in August 1994 and was in full production by April 1995. The fileserver machine is a Sun Sparc 20/612MP, named fnalpubs.fnal.gov. The Sparc station runs the Solaris operating system and includes 256MB of RAM and two 1.05GB hard drives as well as an 8.4GB multidisk pack for a total of 10GB of local disk. The system also uses Andrew File System (AFS) for part of its disk needs. This configuration was recommended by the Operating Systems Support Group.

AFS space is not machine dependent, and allows for easy file access anywhere on the disks from any privileged AFS machine in the world.

**Procedures**

*Delivery Format*

While planning for the pilot program, one of the first decisions the working group had to make was deciding what delivery format to use. The group chose Postscript after studying trends in electronic publishing and meeting with members the SLAC Publications Office, who had been involved in a similar project for two years. (See SLAC trip report). Postscript was already an accepted format for electronic information exchange within the physics community and it also represented the most affordable delivery format in terms of labor, software and hardware costs. Other formats considered were Standardized General Markup Language (SGML), Hypertext Markup Language (HTML) and Portable Document Format (PDF). These formats were rejected for the initial pilot because SGML and HTML were very labor-intensive formats and PDF required an investment in software and was not accessible from machines running VMS and some flavors of UNIX operating systems.

*Workflow*

After Postscript was chosen as the delivery format for the pilot program, the workflow procedures were established. It was decided that reports would be submitted in Postscript format to the technical publications server by the authors. Upon receiving, the Publications Office's technical editor would test the reports using the Ghostview/Ghostscript suite of products to ensure that they could be viewed and printed. The technical editor would then create a LaTeX cover sheet that would be saved as Postscript. The Postscript cover sheet would then be electronically attached to the paper using a script written by Computing Division staff. Finally the paper would be retested to ensure that it was still viewable and printable after the cover sheet was attached. (A cover sheet for each report is required by DOE's Office of Scientific and Technical Information. The cover sheet must contain the Fermilab report number, the Department of Energy disclaimer, distribution authorization, and the Universities Research Association, Incorporated and DOE acknowledgment statement.)

In May 1995, the first technical publications Web pages were published and policies and procedures by which Fermilab authors would transmit their reports to the server were developed and distributed. In June the first full-text preprints in Postscript format were made available on the Internet via the World Wide Web and AFS space. The serving of the first reports successfully ended the pilot program and Phase I of the technical publications fileserver project began.

*Postscript errors*

During Phase I of the project, the procedures for submitting documents worked fairly well. Occasionally files contained Postscript errors that prohibited the technical editor from either attaching the cover sheet, printing or viewing the final document. A Computing Division staff member wrote a script that solved a common Postscript problem found on files generated on Macintosh computers using Microsoft Word. Most of the other problems were unique and had to be handled on a case-by-case basis. During the initial pilot program, the Publications Office and Computing Division staff made an effort to correct as many Postscript errors as staff time allowed. The working group found, however, that correcting many of the errors was too time-consuming and placed too heavy of a burden on the Publications Office, which had only one staff member working full-time on this project. For Phase I of the project, the working group decided that the responsibility for providing a good Postscript file had to ultimately rest with the author. If the Publications Office, with assistance from Computing Division staff, could not easily fix a paper containing a Postscript error, the author was asked to correct the file and resubmit the document. If the author could not fix the file, the paper was distributed in paper format only. During the pilot and Phase I of the project, the Publications Office was able to serve 88% of all submitted reports. (See Figure 1)

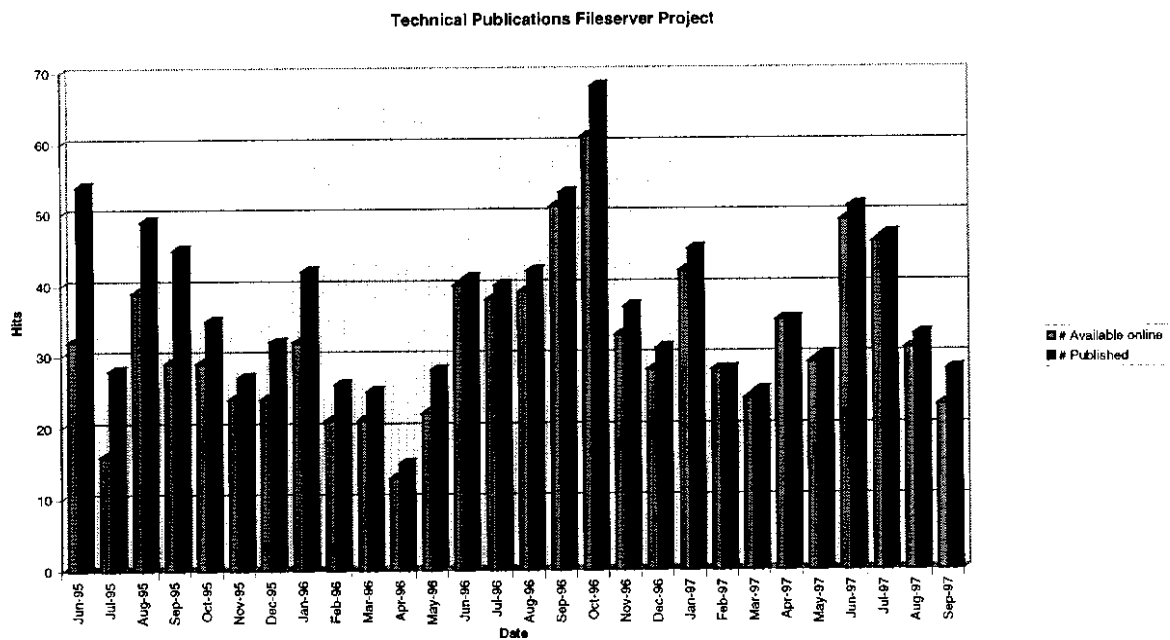

Figure 1

*Organizing directories*

After cover sheets were made and the reports were tested, all technical reports were stored in a consistent directory structure called /archive. Subdirectories were created for each year (e.g., /archive/1996). Further subdirectories represent the type of report (i.e., conference proceeding, physics note, etc.) and then the reports were given file names that corresponded to the report numbers assigned to the reports by the Fermilab Publications Office. For example, a conference report with Fermilab report number FERMILAB-Conf-96/002 resides on the fnalpubs server at: /archive/1996/conf/Conf-96-022.html. The Uniform Resource Locator (URL) for that report is http://fnalpubs.fnal.gov/archive/1996/conf/Conf-96-022.html.

**Expanding Formats**

In 1996 the working group decided to expand the delivery format to include Portable Document Format (PDF) and in 1997 compressed Postscript documents were added. PDF became an option at this time because the Publications Office was able to budget for the software, the technical editor had become proficient at posting papers and was able to add the additional step to her work processes, and the Laboratory moved away from supporting the VMS operating system for which there was no PDF reader. Compressed Postscript was added to enable faster retrieval of online documents and again was possible because of the speed at which the technical editor could process reports.

The following procedures were added: The technical editor creates the PDF files from the author-submitted Postscript documents using Adobe Acrobat Distiller and the compressed Postscript files using gzip. All three formats are then posted to the Web along with file size information.

**Special Services Offered**

As the project progressed, additional customer services were added. Using Web-based forms, customers can request report numbers, order paper copies of reports, and subscribe to an e-mail listserv that notifies subscribers when new reports are posted. Also, the Library created a weekly list on their Web site of new technical reports received in the Library, including Fermilab publications. Links to the full text of the reports are added when available. (See Appendix 1 for related URLs).

**The Library's Role**

Fermilab's technical publications fileserver project dovetailed with the acquisition of a World Wide Web search module for the Library online catalog. The Library's catalog is the search engine for the Laboratory's technical publications and provides hyperlinks to the full-text documents. Users can search the catalog by author, title, call number, keyword, and combination searches.

**Creating the Virtual Library**

In the summer of 1995, the Fermilab Library had the opportunity to beta test Data Research Associate's (DRA) new Web-based interface to their online catalog product. The Library owns the DRA package of integrated library automation products. The new interface module works in tandem with these, layering HTML pages and the Web server over existing Telnet-accessible databases. The Library's automation system runs under the VMS operating system on a MicroVAX 3400 minicomputer.

The Library was very interested in a Web forms-based interface that would rely on industry standards such as Hypertext Transfer Protocol (HTTP), Hypertext Markup Language (HTML) and Z39.50 (the ANSI/NISO Standard for Information Retrieval, Service and Protocols.). The high-energy-physics community was using the World Wide Web for the purpose of sharing physics data and analyses. The Library staff therefore expected that a Web interface for the Library catalog would be quickly embraced by its users.

Because of the Library's experience with the World Wide Web and the high level of sophistication of the online catalog users, the Fermilab Library was able to play an instrumental role in the development of the DRA Web interface module. Weekly bug reports, suggestions and comments were written by the systems librarian and sent to the vendor during the beta test period. Some of the suggestions made were requests for additional formats for record output and ideas for different display parameters.

After a robust test period, the Library received and installed the general release of this software in October 1995. The Library customized the vendor-supplied HTML pages to meet its needs. The Library eliminated many of the graphics supplied by the vendor, rewrote help examples so that they were more appropriate for the Fermilab user community, and tailored the pages to match the Information Resources Department's existing Web pages.

The Web module was Z39.50 and HTTP/HTML compliant and made use of Machine Readable Cataloging (MARC) standards for bibliographic information. The MARC Standard for Electronic Location and Access makes use of a specific field in the bibliographic record (the 856 field) to denote Electronic Location and Access for a particular work. The standard called for the subfield "u" of the 856 field to be used for URLs and allowed the subfield "3" to be used for descriptive labels for the URLs.

The Library had been adding URLs to many of the online catalog bibliographic records since 1994. In 1995, the Library began adding Fermilab technical report URLs. At that time users could copy and paste addresses from the catalog into their graphical Web browsers.

Immediately upon installing the new DRA Web interface, the preprint URLs the Library had been adding to the catalog records became hypertext links to the full-text documents. These direct links made the online catalog more interactive than the previous method of copying and pasting URLs into browsers. Instead of providing only pointers to resources, users could now access the full documents directly from the online catalog.

The Web interface also allowed the Library to add links in the online catalog to electronic journals, Internet information resources, and prepublished physics, engineering and technical papers from other institutions. In this arena, the Library's traditional need to acquire preprints has been diminished, if not eliminated, thereby laying the foundation for a truly virtual library environment.

**Changing Role**

With the new Web interface, the Information Resources Department saw its roles expanding from that of information providers to facilitators of information exchange. Part of the new responsibilities included providing training for customers and making them aware of the tools they need to access technical reports on the Web. Information Resources staff needed to provide tools by which its customers could easily view and print the online documents provided in Postscript and PDF formats. The Ghostview/Ghostscript suite of products and PSTool were recommended for viewing and printing Postscript and Adobe Acrobat Reader was recommended for viewing and printing PDF. These tools were installed on the public access catalog machines in the Fermilab Library. Links on the Information Resources Web pages point to sites where the products can be downloaded.

**Easy Access**

To enhance patron access to the technical publications online collection, the Information Resources Department installed four public access workstations in the Library. The machines are Tektronics XP356 x-stations. The x-terminals are configured with limited access logins, and are set to run the Netscape browser upon startup. The Library welcome page is set as the "home" page. This page includes a large search button link to the online catalog Web interface. All the machines that comprise the Information Resources Department's virtual library system are accessible from one or more of Fermilab's networks, and all of them can connect to the Internet. Currently Ethernet is used for most of the network connections.

## Rapid Growth

Since the technical publications fileserver project began, server usage has increased dramatically. The Information resources Department began gathering usage statistics in October 1996. Fnalpubs usage has grown steadily from 1,062 hits in October 1996 to 66,375 hits in September 1997. (Figure 2).

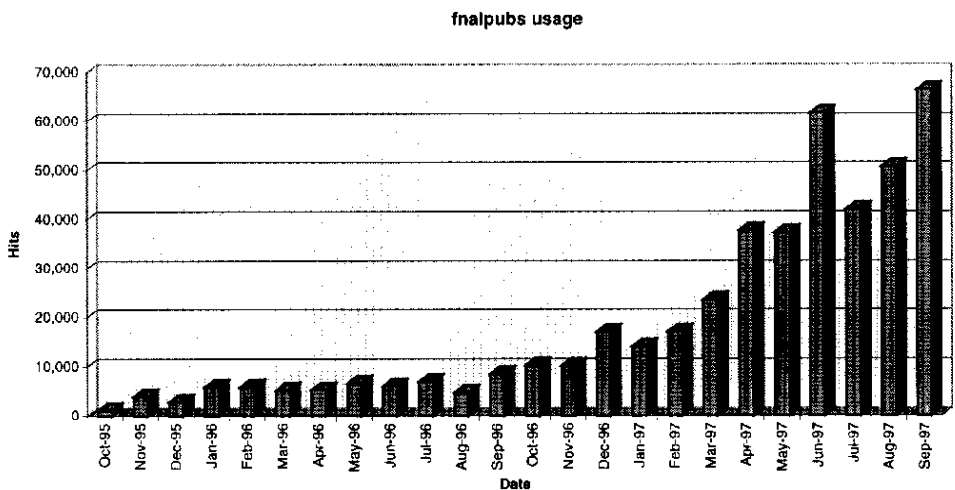

Figure 2

## The Future

Having successfully initiated the technical publications fileserver pilot and Phase I programs, the Technical Publications Working Group is looking to the future. The group has completed a year long study that produced a three-year plan and Phase II recommendations for project expansion.

The recommendations from the working group include developing scripts to automate many of the workflow processes, streamlining the author submission process, a legacy document conversion project, and continued exploration and evaluation of delivery formats that provide easier access, greater searchability and device independence.

The collaboration between the Computing Division and the Information Resources Department has been an excellent cooperative arrangement. The Computing Division provided the technical expertise to make this project successful while the Information Resources Department provided the workflow management, information organization and customer service components. Providing for the information needs of the physics community in the year 2000 and beyond will require the continued cooperation and expertise of both Computing Division and the Information Resources Department.

## Appendix 1 — URLs

Fermilab Information Resources Department
http://fnalpubs.fnal.gov/index.html

Library Home Page
http://fnalpubs.fnal.gov/library/welcome.html

Publications Office Home Page
http://fnalpubs.fnal.gov/techpubs/welcome.html

Library On-line Catalog
http://fnlib.fnal.gov/MARION

Preprint Announcement Mailing List Registration
http://fnalpubs.fnal.gov/techpubs/maillist.html

Fermilab Preprint Number Request Form
http://fnalpubs.fnal.gov/techpubs/numreq.html

Monthly Lists of Fermilab Preprints
http://fnalpubs.fnal.gov/techpubs/pubs_lists.html

Instructions for Electronic Submission
http://fnalpubs.fnal.gov/techpubs/guidelines.html

Processing a Fermilab Technical Report
http://fnalpubs.fnal.gov/techpubs/guideaddendum.html

Feedback
http://fnalpubs.fnal.gov/techpubs/feedback.html